\documentclass{ws-p8-50x6-00}
   \newcommand{\nbi}        {$\rm^{1}$}
   \newcommand{\bnl}        {$\rm^{2}$}
   \newcommand{\newyork}    {$\rm^{3}$}
   \newcommand{\krakow}     {$\rm^{4}$}
   \newcommand{\kraknuc}    {$\rm^{5}$}
   \newcommand{\bergen}     {$\rm^{6}$}
   \newcommand{\ires}       {$\rm^{7}$}
   \newcommand{\texas}      {$\rm^{8}$}
   \newcommand{\oslo}       {$\rm^{9}$}
   \newcommand{\kansas}     {$\rm^{10}$}
   \newcommand{\bucharest}  {$\rm^{11}$}
   \newcommand{\baltimore}  {$\rm^{12}$}

\begin{document}

   \title{Forward Energy and Multiplicity in Au-Au reactions at
   $\sqrt{S_{nn}}=130GeV$}

   \author{Michael Murray}
\address{Cyclotron Institute, Texas A\&M University}
% murray@comp.tamu.edu}                                       \author{
\author{
   I.~G.~Bearden\nbi,
   D.~Beavis\bnl,
   Y.~Blyakhman\newyork,
   J.~Brzychczyk\krakow,
   B.~Budick\newyork,
   H.~B{\O}ggild\nbi,
   C.~Chasman\bnl,
   P.~Christiansen\nbi,
   J.~Cibor\kraknuc,
   R.~Debbe\bnl,
   J.~J.~Gaardh{\O}je\nbi,
   K.~Grotowski\krakow,
   J.~I.~Jordre\bergen,
   F.~Jundt\ires,
   K.~Hagel\texas,
   O.~Hansen\nbi,
   A.~Holm\nbi,
   C.~Holm\nbi,
   A.~K.~Holme\oslo,
   H.~Ito\kansas,
   E.~Jacobsen\nbi,
   A.~Jipa\bucharest,
   C.~E.~J{\O}rgensen\nbi,
   E.~J.~Kim\baltimore,
   T.~Kosic\krakow,
   T.~Keutgen\texas,
   T.~M.~Larsen\oslo,
   J.~H.~Lee\bnl,
   Y.~K.~Lee\baltimore,
   G.~L{\O}vhoiden\oslo,
   Z.~Majka\krakow,
   A.~Makeev\texas,
   B.~McBreen\bnl,
   M.~Murray\texas,
   J.~Natowitz\texas,
   B.~S.~Nielsen\nbi,
   K.~Olchanski\bnl,
   D.~Ouerdane\nbi,
   J.~Olness\bnl,
   R.~Planeta\krakow,
   F.~Rami\ires,
   D.~Roehrich\bergen,
   B.~Samset\oslo,
   S.~Sanders\kansas,
   R.~A.~Sheetz\bnl,
   P.~Staczel\nbi,
   T.~F.~Thorsteinsen\bergen$^+$,
   T.~S.~Tveter\oslo,
   F.~Videb{\AE}k\bnl,
   R.~Wada\texas,
   F.~Rami\ires,
   A.~Wieloch\krakow,
   I.~S.~Zgura\bucharest \\
   the BRAHMS Collaboration
   }
%   Z.~Sosin\bucharest,

   \address{
   \nbi~Niels Bohr Institute,
   \bnl~Brookhaven National Laboratory,
   \newyork~New York University,
   \krakow~Jagiellonian University, Krakow,
   \kraknuc~Institute of Nuclear Physics, Krakow
   \bergen~University of Bergen,
   \ires~IReS and Universit\'e Louis Pasteur, Strasbourg
   \texas~Texas A$\&$M University,
   \kansas~University of Kansas,
   \oslo~Fysisk Institutt, University of Oslo,
   \bucharest~University of Bucharest,
   \baltimore~John Hopkins University,
   $^+ Deceased$
   }


   \maketitle

\abstracts{
For relativistic heavy ion collisions the energy flow in the collision reveals
information on the equation of state of matter at high density.  The BRAHMS
experiment has studied the relation between multiplicity in the side-wards
direction and zero degree  energy using our Tile Multiplicity Array and Zero
Degree Calorimeters. To understand this spectrum requires a knowledge of the
coalescence of nucleons that are close in phase space. We have also studied
electromagnetic collisions and compared our results to lower energy data. As the
center of mass energy increases, each nucleus sees a stronger electromagnetic
field  resulting in more energy being absorbed. This in turn causes an increase
in the neutron multiplicity.
}
   \section{The Experimental Setup}
   The BRAHMS experiment consists of two movable spectrometers and four
   detector systems to measure  global   variables for each event,\cite{BRAHMS}.
   The spectrometers allow BRAHMS to measure particles yields over a very wide
   range of $p_T$ and rapidity. In this paper only details of the global detectors used in
   the present analysis are given.

   The Tile Multiplicity Array, TMA, consists of an
   hexagonal barrel 96cm long placed around the beryllium beam pipe. The array is
   made up of square tiles that are 12cm long and 0.5cm thick.   For collisions at
   the
   nominal interaction point the pseudo-rapidity range is $|\eta|<2.2$.
   Four tiles in front of each of the two spectrometers have been removed to
   minimize
   multiple scattering of particles heading into the spectrometers.  The
   multiplicity is
   measured by counting the energy deposited in each ring and dividing by the
   mean energy lost per particle.

   The Beam-Beam counter consists of two sparsely populated arrays of
    Cerenkovs (made from Ultra Violet Transmission plastic)
   coupled to
   photomutiplier tubes. They are $\pm2.3$m from the
   nominal interaction vertex and subtend the pseudo-rapidity regions
   $3.0<|\eta|<3.5$. The left array has eight 2 inch tubes
   and 36 3/4 inch ones arranged symmetrically  around the beam pipe.
   In order to allow the forward spectrometer to move to small angles
   the right array is asymmetric. It consists of 30 small tubes and 5 large
   ones.

   Finally two  zero degree calorimeters (ZDCs) at $\pm 18$m measure the energy
   of forward going neutral particles\cite{ZDCNIM}. Almost all of this energy is
   from
   non interacting neutrons. They are used for finding the
   interaction vertex and for centrality determination. The ZDCs are also very useful
   to
   study peripheral electromagnetic interactions and provide a measure of the
luminosity for the accelerator and experiment,\cite{BALTZ}.

   \section{Forward Energy versus Side-wards Multiplicity}
\begin{figure}[t]
%   \figurebox{23pc}{15pc}{} % to have a box alone
   \epsfxsize=23pc
   \epsfbox{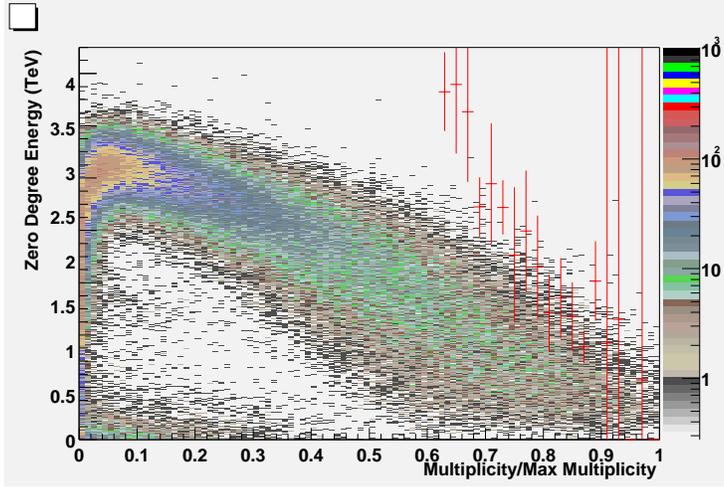}
   \caption{ZDC total energy versus multiplicity in the tile counters. The overlaid
   histogram represents the prediction of the AMPT \protect\cite{AMPT} model.
   \label{fg:zdcvmult}}
   \end{figure}

   One of the most import things to know about heavy ion collisions is the number
   of
   nucleons that participated in the collision, $n_{participants}$. The relationship
   between
   the number of participants and the multiplicity and central rapidity is a measure
   of
   how much energy each nucleon loses in the collision and how this energy is
   shared
   between particle production and particle motion.
   This is best  done by measuring the forward   going energy  of those nucleons
   that
   do not participate using a zero degree calorimeter,\cite{NA49spect}.
   Unfortunately at RHIC the
   beam focusing magnets sweep all charged particles away from the ZDCs and only
   the neutrons are measured. This complicates the extraction of the centrality of
   the
   collision from the ZDC measurement.
   %sigma(b) goes increases 15 (n+p) to 19% (only n).
   Figure~\ref{fg:zdcvmult} shows the correlation between the forward going neutral
   energy and the multiplicity for $|\eta|<2.$ as measured in the tiles.  For
   events in
   the top 70$\%$ of the multiplicity distribution there is a steady decrease of
   neutron multiplicity with energy. While there is certainly a strong correlation
   between the  ZDC  energy and the  multiplicity at mid-rapidity the distribution
   is
   wide. This is due partly to the small number of neutrons observed and to the detectors'
   resolution. However it seems likely that intrinsic fluctuations between the number of
   participants and the  multiplicity at mid-rapidity  are also important.

    The overlaid
   histogram represents the prediction of the AMPT \protect\cite{AMPT} model.
   Both AMPT and HIJING\cite{HIJING}, overestimate the number of forward going neutrons.
   This may be because it ignores
   the possibility that  nucleons   can coalesce into light clusters such as
   deuterons or
   tritons. Any coalescence effect would decrease the number of forward going neutrons.
      The JAM code includes
   coalescence of nucleons that are close together in phase space\cite{JAM}. This
   does a much  better
   job of reproducing the ZDC spectrum.

   \section{Electromagnetic Interactions}
     Peripheral interactions of heavy ions produce intense electromagnetic fields
   which
   may be used to probe the structure of the nucleus. At $\sqrt{S_{nn}}=130GeV$
   one nucleus sees the other approaching at a speed corresponding to a Lorenz
   factor
   of $\gamma=9800$. The nucleus may absorb several photons and be excited into
   the giant dipole or even to multi-phonon resonances. As the nucleus equilibrates and
   cools
   it may emit several neutrons, protons and photons.

\begin{figure}[t]
%   \figurebox{23pc}{15pc}{} % to have a box alone
   \epsfxsize=23pc
   \epsfbox{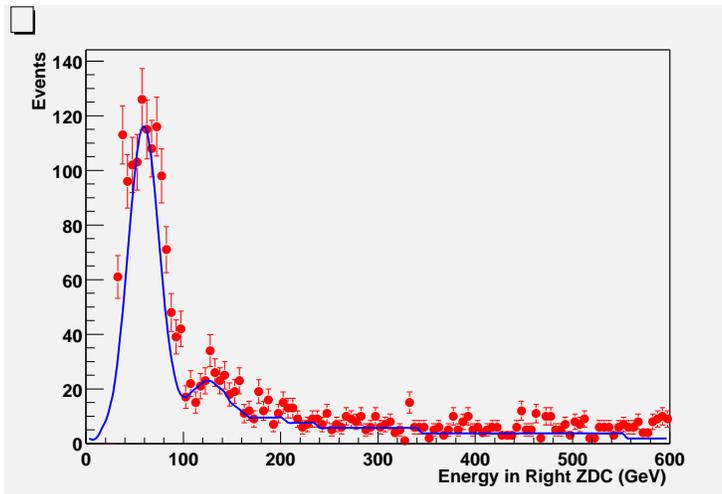}
   \caption{The Zero Degree Calorimeter spectrum for electromagnetic events.
The smooth curve shows the prediction of the RELDIS model.
   \label{fg:emdis}}
   \end{figure}

     In order to study electromagnetic interactions we
    require no hits in either the tiles or the beam-beam counters.
   From HIJING\cite{HIJING} we estimate that we have
   a $2\%$   contamination of peripheral nuclear events in the sample.
   Figure~\ref{fg:emdis} shows the energy spectrum in the right ZDC.
   The spectrum shows a strong one neutron peak, a smaller two neutron peak and
   then a continuum.   Figure~\ref{fg:emdis} also shows
   the energy spectrum from the  RELDIS\cite{RELDIS} model after smearing with the
   experimental resolution.
   RELDIS  extends the
    Weizsacker-Williams method to Mutual Coloumb Dissociation.
        Photo-neutron cross sections measured in different experiments
   are used as input for the calculations of dissociation cross sections.
    The excited nuclei reach thermal equilibrium and then decay
   according to the statistical
   evaporation-fission-multi-fragmentation model, (the SMM model)~\cite{JPB}.
   The data and model have been normalized at the one neutron peak. The real
   spectrum
   is
   consistently higher than the spectrum from the model.

\begin{figure}[t]
   \epsfxsize=23pc
   \epsfbox{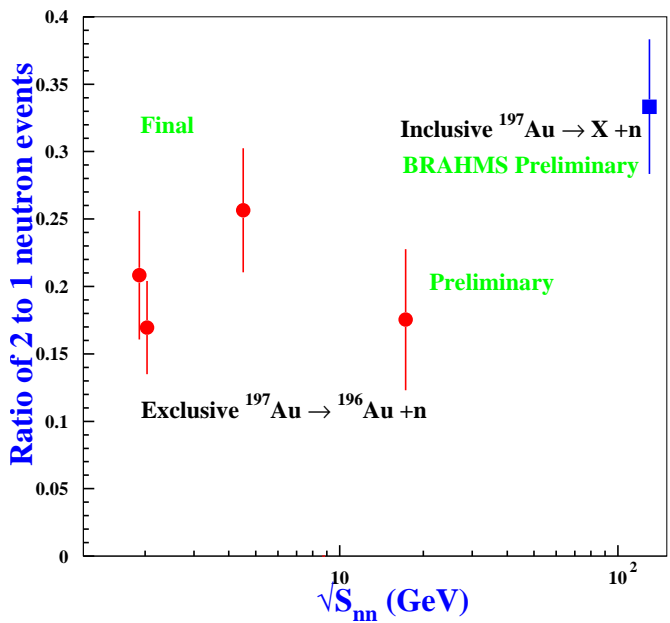}
   \caption{Ratio of two to one neutron cross sections as a function of
   $\sqrt{s_{nn}}$. The data below $\sqrt{s_{nn}}=130GeV$ are from
   \protect\cite{hill}.
   \label{fg:r21ve}}
   \end{figure}

   Since only non-interacting nucleons hit the ZDCs their energy spread is set by
   their Fermi motion  in the nucleus. This is negligible compared to the energy
   resolution
   of the ZDC which is $\sigma_E=14GeV$ for one $65GeV$ neutron. Therefore we can
   fit the spectrum to a series of Gaussians at $n\cdot 65GeV$ with width
   $\sqrt{n} \cdot \sigma_E$ and extract the multiplicity of 1, 2, 3 etc neutrons.
    Figure~\ref{fg:r21ve} shows the ratio of two to one neutron cross sections as a
   function of $\sqrt{s_{nn}}$.  The probability  to emit 2 neutrons rises with
       $\sqrt{s_{nn}}$, because more energy is absorbed and the
   multi-phonon  resonances are more likely to be excited.   Such a correlation
   between
   the excitation energy of a compound nucleus and the neutron multiplicity has
   been observed
   in $CuAu$ collisions at % beam energy of
   35A MeV,\cite{Wada97a}.

   \section{Conclusions}
       For nuclear collisions the number of neutrons emitted forward is anti-correlated
to the multiplicity observed at central pseudo-rapidity. However the distribution is
rather broad. In order to reproduce the number of neutrons observed it is necessary
to estimate the number that are removed from the spectrum by coalescence.
For electromagnetic collisions the RELDIS model gives a reasonable
description of the neutron spectrum at zero  degrees. In this model the
electromagnetic breakup of
two nuclei is mainly due to multi-photon exchange. As the beam energy increases the
neutron multiplicity in electromagnetic interactions increases.

   \section*{Acknowledgments}
This analysis would not have been possible without the  very successful commissioning
of RHIC by the C-AD department of BNL.
   I would like to thank the organizers of ISMD for their invitation and generous
   support.
This work
   was supported in part by the DOE under contract FG03-93-ER40773.
 by the Danish Natural Science
Research Council, by the Norwegian Natural Science Research Council and by
the Polish ministry of Research.

   \end{document}